# Deciphering Spatial and Multi-scale Variations in the Effects of Key Factors of Maritime Safety: A Multi-scale Geographically Weighted Approach

Guorong Li[a,b], Kun Gao[b,*], Jinxian Weng[a], Xiaobo Qu[b]

[a] College of Transport and Communications, Shanghai Maritime University, Shanghai, 201306, China.

[b] Department of Architecture and Civil Engineering, Chalmers University of Technology, Gothenburg, SE-412 96, Sweden.

Email: gkun@chalmers.se

**Abstract**

Maritime accidents and corresponding consequences vary substantially across spatial dimensions as affected by various factors. Understanding the effects of key factors on maritime accident consequence would be of great benefit to prevent the occurrence or reduce the consequences of maritime accidents. Based on unique maritime accident data with geographical information covering fifteen years in the East China Sea, a multi-scale geographically weighted regression (MGWR) model considering the multi-scale spatial variation is employed to quantify the influences of different factors as well as the spatial heterogeneity in the effects of key factors on maritime accident consequence. The performances of MGWR are compared with multiple linear regression (MLR) and geographically weighted regression (GWR). Especially, MGWR outperforms the other two models in terms of modeling fitness and clearly capturing the unobserved spatial heterogeneity in effects of factors. Results reveal notably distinct and even inverse influences of some factors in different water areas on maritime accident consequences. For instance, approximately 50% of the accident locations present positive coefficients of good visibility while other locations are negative, which are ignored by MLR. The outcomes provide insights for making appropriate safety countermeasures and policies customized for different geographic areas.

# 1. Introduction

The shipping industry has been a catalyst for economic development and prosperity. Approximately 80 percent of global trade volume is seaborne (UNCTAD, 2020). The increment of maritime transportation, on the one hand, could accelerate the production of air pollutants (Shi et al., 2020; Wang et al., 2021b). On the other hand, high maritime traffic volume also leads to a higher probability of marine accidents, which might cause catastrophic consequences such as property damage, fatality, and environmental pollution (Goerlandt and Kujala, 2011). European Maritime Safety Agency (2020) reported that there had been 21,392 ships involved in 19,418 casualties/incidents, with 496 fatalities and 6210 persons injured from 2014 to 2019. Maritime safety has attracted much attention in recent decades and how to enhance maritime safety remains a great concern of both maritime logistic operators and maritime managers (Eliopoulou and Papanikolaou, 2007). In order to effectively reduce the occurrence of maritime accidents and associated consequences, one of the cores is to interrogate and quantify the key factors of maritime accident rates and the severity of accidental consequences. These understandings and knowledge could support corresponding and effective countermeasures to reduce hazards in the future (IMO, 2008; Li et al., 2012; Luo and Shin, 2019). These motivate both academic researchers and practitioners to analyze the impacts of various factors on maritime accident consequences.

Previous studies have investigated maritime accidents and consequences from different aspects, such as casualties and loss of life (e.g., Roberts et al., 2013), property damage cost (e.g., Weng et al., 2015), oil spill (e.g., Talley et al., 2012), and accident severities (e.g., Talley et al., 2008; Wang et al., 2021a). Meanwhile, many approaches have been proposed to quantify the impacts of different factors on the investigated dependent variable, to name a few, fuzzy matter element methods (e.g., Chen et al., 2017), Poisson regression model (e.g., Yip et al., 2015), discrete choice model (e.g., Jin, 2014; Shi et al., 2021), zero-inflated models (e.g., Chai et al., 2018). However, the studies mentioned above generally investigated maritime accidents from a global perspective. In recent years, some other studies have started to focus on the spatial patterns of maritime accidents (e.g., Huang et al., 2013; Zhou et al., 2020; Rong et al., 2021; Zhang et al., 2021; Liu et al., 2022). For instance, Zhou et al. (2020) mapped the maritime accident risk of the South China Sea, and Zhang et al. (2021) studied the spatial patterns and characteristics of global maritime accidents. These studies imply that maritime accidents vary in spatial dimensions, but did not further investigate the influencing factors of the spatial patterns of maritime accidents. Namely, the influencing factors and their effect degree on maritime accidents are not comprehensively and quantitatively investigated based on adequate data. More importantly, the potential spatial heterogeneity in the effect of a factor on accident consequence is seldom explored. The spatial heterogeneity in the effect of a factor refers to distinct effects of a factor on the investigated dependent variable in different geographical locations (Fotheringham et al., 2002; Gao et al., 2021). For instance, the wind may have different effects (e.g., degree of effect) on maritime accident consequences in different ocean areas. The potential spatial heterogeneity in the effects of a factor on maritime accident consequence matters in formulating safety improvement strategies. For a specific instance, if the strong wind has distinct degrees of effects on maritime accident consequences in different water areas, different thresholds should be used to determine risk warning or class in different areas, and more strong interventions should be adopted in areas where the strong wind has more notable effects. Hence, the existing limitations may result in inadequate understanding concerning significant factors affecting maritime accident consequences, and thus inefficient countermeasures to improve maritime system safety in different areas.

Standing on the wake of existing studies, this study aims to investigate and quantify the influences of various factors on maritime accident consequences with specific considerations of spatial heterogeneity. This targets to explore effective improvement countermeasures for the safety of maritime transportation systems. A dataset including 1248 maritime accident records and relevant environment information in the East China Sea from 2000 to 2014 is used for analysis. The spatial pattern of the maritime accident is firstly analyzed using spatial Kernel Density Estimation (KDE). More importantly, the multi-scale geographically weighted regression (MGWR) is further utilized to investigate the key factors influencing maritime accident consequences (measured by overall economic losses considering all aspects) and the potential distinct effects of a factor in different water areas (i.e., spatial heterogeneity) with consideration of discrepancies in the spatial scales. To the best of our knowledge, this is the first attempt to investigate the spatial heterogeneity in the effects of various factors on maritime accident consequences based on adequate data. The results contribute to the comprehension of the maritime accident consequences from a spatial perspective and are beneficial for excavating practical implications. More specifically, results are helpful in establishing differential strategies in facilitating maritime safety in different water areas by preventing serious accidents, as well as providing hints for other applications such as designing insurance rates for maritime accidents in different water areas.

The remainder of this paper is organized as follows. Section 2 reviews the relevant literature to highlight the research gaps and the contributions of this study. Section 3 describes the used dataset in terms of study water area and data contents. Section 4 introduces the methodologies for analysis. Section 5 and 6 present the model results and discussions, respectively. Conclusions and future works are summarized in the last section.

## 2. Literature review

To reduce maritime accidents (e.g., occurrence likelihood and corresponding consequences), studies have been dedicated to analyzing maritime accidents from different angles and exploring potential implications for effective measures based on various data resources such as automatic identification system (AIS) data, simulation data, and maritime accident data (e.g., Szlapczynski et al., 2021; Cai et al., 2021; Mazurek et al., 2022; Gil et al., 2022; Ma et al., 2022). Among them, AIS data and simulation data are generally used in collision avoidance analysis and collision frequency estimation. Just to name a few, Mou et al. (2010) presented a collision-avoidance analysis in the busy water area of Rotterdam Port. With the collected AIS data, this paper firstly identified the correlation between the closest point of approach and influencing factors, including ship size, ship speed, and ship course. A dynamic methodology based on SAMSON was subsequently proposed to assess the real-time collision risk. Altan and Otay (2018) developed a collision model based on molecular collision theory and long-term AIS data as inputs to estimate and visualize the encounter probability in congested waterways. Ship density, relative velocity, and collision diameter were considered to be the critical factors that affected the encounter probability. Similarly, Szlapczynski et al. (2021) developed a near-miss detection method for collision alert systems based on ship domain theory and verified the model with simulation data from three scenarios. Using AIS data as the input, Murray and Perera (2021) presented a deep learning framework to predict regional ship behavior. These studies mainly focused on evaluation methods and developing collision-avoidance systems and strategies for maritime accident prevention.

Another stream of studies is to leverage maritime accident databases for analysis (Wang and Yang, 2018; Çakır et al., 2021a, Çakır et al., 2021b). Various studies have been carried out to evaluate injury/loss of human life or property damage cost for different ship types and water areas, investigate influencing factors and thus seek efficient measures for preventing similar accidents from occurring in the future. Jin (2014) estimated vessel damage severity and crew injury severity of fishing vessel accidents using the ordered probit model. Seven types of explanatory variables, including accident type, vessel characteristics, propulsion type, hull construction type, weather condition, spatial information, and time of accidents, were considered in this paper. Yip et al. (2015) explored the determinants of injuries in passenger vessel accidents using Poisson regressions and empirical data on ferry, ocean cruise, and river cruise vessel accidents from the U.S. Coast Guard. The model took into account the majority of influencing factors as well. The results indicated that the number of passenger injuries was positively related to the number of crew injuries. Weng et al. (2018) developed a generalized F-distribution model with random parameters to estimate the property damage costs in maritime accidents. Random parameters were considered to represent the changeable effects of variables on different observations. Four types of variables, including accident characteristics, ship types, environmental characteristics, and human causal factors, were considered in their study. Based on the worldwide accident investigation reports from 2010 to 2019, Wang et al. (2021) applied an ordered logistic regression model to reflect the relationship between different factors and the severity of marine accidents. Furthermore, Çakır et al. (2021a) utilized various association rule mining algorithms to investigate the factors affecting tugboat accidents by analyzing the tugboat accident dataset from the Information Handling Services (IHS) Sea-Web database. It was found that serious maritime accidents were related to hull/machinery damage.

Although the aforementioned studies have investigated the relationship between some influencing factors and consequences of maritime accidents such as human injury/loss and property damage cost, they have only focused on modeling partial factors due to data limitations and a part of the accident consequences (e.g., only estimate fatality loss or property damage cost). Moreover, the modeling methods in existing literature generally assume the influences of a factor are global in different geographical locations, and neglect the potential spatial heterogeneity in the effects of an influencing factor on maritime accident consequences. The effects of some influencing factors may show distinct effects on the maritime accident consequence in different water areas. Investigation regarding the potential spatial heterogeneity in effects of factors is important for making tailored safety improvement measures for different areas. For instance, if a factor has much more noticeable impacts in some water areas, more intense regulations for alleviating the effects of this factor should be imposed to ensure safety in such areas as compared to other water areas. However, the potential spatial heterogeneity in the effects of factors is overlooked in the majority of prior studies and has never been explored in the area of maritime accident analysis, to our best knowledge. Although Jin (2014) had considered the spatial information during the model evaluation, they only added a spatial variable (distance to shore) in the model formulation, which can hardly model the spatial heterogeneity in the effects of a factor on maritime accidents.

Benefiting from advances in information and communication technology, maritime accident data with geographic information become available from data sources such as real-time AIS records of the ships involved in accidents. Such data give researchers the opportunity to explore the potential spatial heterogeneity in the effects of various factors on maritime accident consequences. As far as we are concerned, no existing work has investigated the spatial heterogeneity in the effects of key factors of maritime accident consequence based on long-period maritime accident data. Therefore, this study endeavors to reveal the key factors of maritime accident consequences. Especially, we take into consideration of spatial heterogeneity in the effects of influencing factors in modeling and analysis to explore the distinguished impacts of a factor on maritime safety in different geographical areas. These are salutary for making tailored safety improvement strategies and measures in different water areas, which eventually help to prevent maritime accidents and to expedite the safety of maritime transportation.

## 3. Data description

The study area is Fujian Province, an important coastal province in the East China Sea. Fujian contains a large jurisdictional water area of 136 thousand $km^2$ (see Fig. 1). It takes an important position in external trades in China and covers vital shipping routes between the South China Sea and the East China Sea. The coastline in Fujian is 3752 kilometers long, with the highest coastal meander rate (1:7.01) in China. Hundreds of bays and countless islands make the navigational environment complicated in the Fujian water area, which leads to potential risks of maritime accidents. The waters are mainly managed by the authorities of six major coastal cities, including Ningde, Fuzhou, Putian, Quanzhou, Xiamen, and Zhangzhou (Fig. 1).

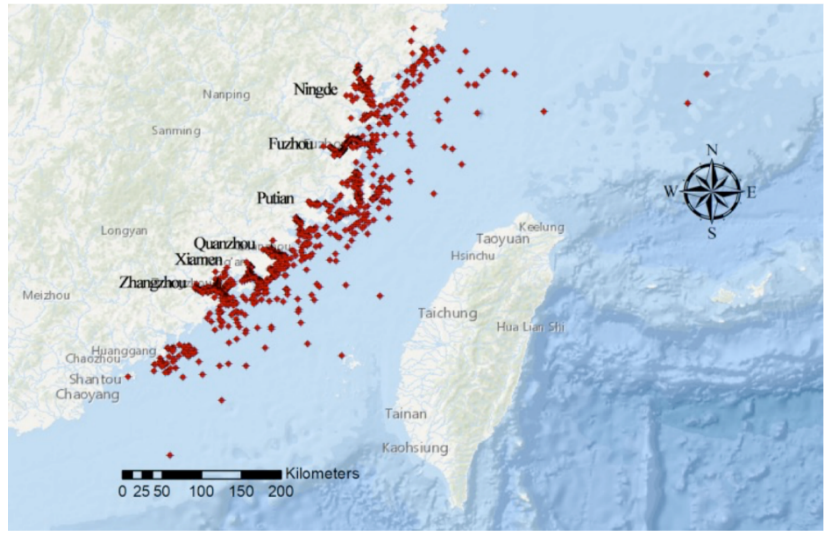

**Fig. 1.** The study water area and locations of recorded maritime accidents.

This study utilizes maritime accident data in the study area provided by Fujian Maritime Safety Administration (FSA). The dataset contains 1248 maritime accidents with longitude and latitude information recorded from 2000 to 2014, as shown in Fig.1. The large quantity of data highlights the uniqueness of the data and guarantees the representativeness of exploring significant factors of maritime accidents. The maritime accident consequence, the dependent variable in this study, refers to the overall economic cost of the accident reported by the authority considering all aspects such as injured or death/missing crews and property damage cost. In the dataset, the economic loss from maritime accidents ranged from 2,000 to 32 million CNY (1 CNY ≈ 0.137 USD). Moreover, the dataset also includes various potential factors that may influence accident consequences and can be summarized into four categories, including accident types, ship information, environmental characteristics, and reported accident causes. The reported accident causes refer to the reported causes that potentially led to the accidents in the accident report recorded by the authorities. The statistics about these factors are summarized in Table 1. More specifically, the dataset covers seven types of accidents. Ship information consists of types, gross tonnage, and navigational status of the ships. Environmental characteristics contain accident time of the day, visibility condition, and strong wind/wave or typhoon. Reported accident causes include judgment error, lookout failure, misoperation, and machinery failure refer to the common factors leading to accidents. Table 1 tabulates the detailed statistical descriptions concerning the influencing factors mentioned above. It is worth clarifying that many explanatory variables in our dataset are coded as categorical variables rather than continuous variables. This is a common case in the relevant literature of using accident datasets (Weng et al., 2018; Wang and Yang, 2018; Çakır et al., 2021a; Çakır et al., 2021b). The reason is that those are all available information complied in accident reports by the management authority, who did not record exact values of some factors such as the value of wind speed during the accident. The dependent variable (accident consequence) and the factor (gross tonnage) are continuous variable and standardized to eliminate the effects of different units of the two factors.

**Table 1. Descriptive statistics of the maritime accident characteristics in the Fujian water area.**

| Variables | Descriptions | Types | Moran's I |
|---|---|---|---|
| **Dependent variable** | | | |
| Accident consequence | The human life loss and property damage cost resulting from maritime accidents ×$10^4$ CNY (99.405) | Continuous | 0.047 |
| **Accident types** | | | |
| Collision | 1 = collision (41.3%), 0 = otherwise (58.7%) | Categorical | 0.418* |
| Contact | 1 = contact (17.1%), 0 = otherwise (82.9%) | Categorical | 0.559* |
| Striking rock | 1 = striking rock (10.8%), 0 = otherwise (89.2%) | Categorical | 0.161 |
| Grounding | 1 = grounding (10.4%), 0 = otherwise (89.6%) | Categorical | -0.053 |
| Sinking/capsizing | 1 = sinking/capsizing (7.4%), 0 = otherwise (92.6%) | Categorical | 0.377* |
| Fire/explosion | 1 = fire/explosion (4.5%), 0 = otherwise (95.5%) | Categorical | 0.027 |
| Other accident types | 1 = other accident types (4.9%), 0 = otherwise (95.1%) | Categorical | -0.001 |
| **Ship information** | | | |
| Two dry cargo ship involved | 1 = two involved (16.9%), 0 = otherwise (83.1%) | Categorical | 0.282* |
| One dry cargo ship involved | 1 = one involved (66.7%), 0 = otherwise (33.3%) | Categorical | 0.126 |
| Fishing ship | 1 = involved (14.9%), 0 = otherwise (85.1%) | Categorical | 0.208 |
| Liquid cargo ship | 1 = involved (7.4%), 0 = otherwise (92.6%) | Categorical | 0.063 |
| Other ship types | 1 = involved (17.9%), 0 = otherwise (82.1%) | Categorical | 0.482* |
| Gross tonnage | The sum of tonnage for ships involved in the accident (5699.626) | Continuous | 0.257 |
| Navigational status | 1 = underway (82.2%), 0 = moored/docked (17.8%) | Categorical | 0.426* |
| **Environmental characteristics** | | | |
| Time of the day | 1 = nighttime period (50.6%), 0 = daytime period (49.4%) | Categorical | 0.245* |
| Good visibility | 1 = yes (17.4%), 0 = no (82.6%) | Categorical | 0.181 |
| Restricted visibility | 1 = yes (6.7%), 0 = no (93.3%) | Categorical | 0.015 |
| Strong wind/wave | 1 = yes (25.5%), 0 = no (74.5%) | Categorical | 0.096 |
| Typhoon | 1 = yes (4.1%), 0 = no (95.9%) | Categorical | 0.032 |
| **Accident cause factors** | | | |
| Judgment error | 1 = yes (10.5%), 0 = no (89.5%) | Categorical | -0.080 |
| Lookout failure | 1 = yes (34.9%), 0 = no (65.1%) | Categorical | 0.135 |
| Operation error | 1 = yes (55.1%), 0 = no (44.9%) | Categorical | 0.011 |
| Machinery failure | 1 = yes (7.8%), 0 = no (92.2%) | Categorical | 0.352* |

Note: The statistics in parentheses: mean values for continuous variables, percentages for categorical variables; Average exchange rate during 2000-2014: 1 CNY=0.13732 USD; *$p \leq 0.05$.

## 4. Methodology

### 4.1 Spatial correlation analysis

For investigating spatial heterogeneity in the effects of influencing factors, it is vital to first check the potential correlations in spatial dimensions (Fotheringham et al., 2002; Gao et al., 2021; Çakır et al., 2021a). More specifically, if a factor does not present spatial correlations at all, it is not necessary to use methods like GWR to investigate the influences of different factors on maritime accident consequences. This study selected the global Moran's I index to measure the potential spatial autocorrelation of a factor. Moran's I coefficient (Moran, 1950) is one of the most widely used measures for detecting spatial autocorrelation. The equation for the Moran's I is

$$I = \frac{n}{\sum_{p=1}^{n}\sum_{q=1}^{n}w_{pq}} \times \frac{\sum_{p=1}^{n}\sum_{q=1}^{n}w_{pq}(x_p - \bar{x})(x_q - \bar{x})}{\sum_{p=1}^{n}(x_p - \bar{x})^2} \quad (1)$$

where $n$ is the number of observations, $x$ represents the variable of interest, $\bar{x}$ is the average value of the variable, $p$ and $q$ denote the location indices, $w_{pq}$ represents the matrix of spatial weights given by a selected geographical criterion with diagonal elements equal to zero (i.e., $w_{pp} = 0$). The weight matrix used in this study is mentioned in subsection 4.3. The value of the Moran' I index ranged from – 1 to + 1. A larger value indicates a higher degree of spatial clustering, while a lower value indicates a higher degree of spatial dispersion. We calculated the global Moran's I index for all the variables, as shown in Table 1. Eight of the variables show significant spatial correlation with 95% confidence levels. Other variables present less obvious spatial relationships or less significant spatial correlations. The large differences of the Moran's I index among different variables indicate that there may exist significant variances in spatial correlation or dispersion of these variables. As far as we are concerned, the degree to which variables are aggregate or discrete may affect bandwidth selection during the GWR modeling process. Therefore, it is reasonable to consider bandwidth variations in analysis to obtain more accurate modeling using MGWR.

*4.2 Multiple Linear regression*

As a baseline method, a multiple linear regression (MLR) is used preliminarily to model the relationship between the maritime accident consequence and available explanatory factors. The aim is to preliminarily check the important factors of maritime accident consequences. The MLR model is formulated as

$$y_i = \beta_0 + \sum_{j=1}^{k}\beta_j x_{ij} + \varepsilon_i \quad (2)$$

where $y_i$ represents the consequence of maritime accident $i$ and is measured by the overall economic loss of the accident; $x_{ij}$ is the $j^{th}$ explanatory variable of accident $i$, $\beta_j$ stands for the corresponding coefficient of the $j^{th}$ explanatory variable while $\beta_0$ is intercept term; $\varepsilon_i$ is the random error term and follows normal distributions with an average of zero. As indicated by Eq (2), MLR assumes the effect of a factor to be global across all data points in the dataset. Therefore, the estimated coefficient of a factor reflects the globally average effect of a factor on maritime accident consequences.

*4.3 Multi-scale geographically weighted regression*

Geographically Weighted Regression (GWR), as a spatial regression technique, is widely employed to investigate spatially nonstationary relationships (Fotheringham et al., 2002; Ziakopoulos and Yannis, 2020; Gao et al., 2021). Leveraging the locally weighted least square method, GWR is capable of incorporating location information of the samples into the modeling process so that it can produce exclusive coefficients for different sample points. This enables GWR to consider the spatial effects of independent variables and analyze the differences in the effects of a factor across different locations (i.e., spatial heterogeneity in the effects of a factor). The formulation of GWR is

$$y_i = \beta_0(u_i, v_i) + \sum_{j=1}^{k} \beta_j(u_i, v_i)x_{ij} + \varepsilon_i \tag{3}$$

where $(u_i, v_i)$ represents the location of the accident $i$. Other coefficients have same meanings as they are in Eq. (2). The locally weighted least square minimization is used to estimate $\beta_j(u_i, v_i)$ based on available data and can be expressed as

$$\min \sum_{i=1}^{n} w_i(u_i, v_i)[y_i(u_i, v_i) - \sum_{j=1}^{k} \beta_j(u_i, v_i)x_{ij}]^2 \tag{4}$$

where $w_i(u_i, v_i)$ denotes the spatial weight at location $(u_i, v_i)$. The coefficients can be estimated by

$$\hat{\beta}(u_i, v_i) = \frac{X^T W(u_i, v_i) Y}{X^T W(u_i, v_i) X}$$

$$X = \begin{bmatrix} x_{01}, x_{11}, ..., x_{k1} \\ x_{02}, x_{12}, ..., x_{k2} \\ ... \quad ... \quad ... \quad ... \\ x_{0n}, x_{1n}, ..., x_{kn} \end{bmatrix}; \quad Y = \begin{bmatrix} y_1 \\ y_2 \\ ... \\ y_n \end{bmatrix}; \quad W(u_i, v_i) = \begin{bmatrix} w_1(u_i, v_i), 0, ..., 0 \\ 0, w_2(u_i, v_i), ..., 0 \\ ... \quad ... \quad ... \quad ... \\ 0, 0, ..., w_n(u_i, v_i) \end{bmatrix} \tag{5}$$

The location-specific coefficient $\hat{\beta}(u_i, v_i)$ allows the relationship between a factor and maritime accident consequences to vary across different water areas. More clearly, a specific coefficient of a factor is estimated for each data point (Ziakopoulos and Yannis, 2020; Gao et al., 2021). One key component in Eq. (5) is determining the spatial weight matrix, which is generally negatively related to the distance between locations. We adopt an adaptive bi-square kernel function as the distance-weighting function referring to literature (Fotheringham and Oshan, 2016).

$$w_{pq} = \begin{cases} \left[1 - (\frac{d_{pq}}{b})^2\right]^2 & \text{if } d_{pq} < b \\ 0 & \text{otherwise} \end{cases} \tag{6}$$

where $w_{pq}$ represents the weight between accident location $p$ and accident location $q$; $d_{pq}$ is the distance between $p$ and $q$; $b$ is a critical distance from location $p$ to its $M^{th}$ nearest neighbor. $M$ is the optimal number of nearest neighbors, namely the bandwidth of the model, which is determined by minimizing the corrected Akaike Information Criterion (AICc). A larger bandwidth implies a smaller spatial heterogeneity (Fotheringham et al., 2002). The goodness-of-fit measure AICc is defined by

$$AICc = 2n\ln(\hat{\sigma}) + n\ln(2\pi) + n\frac{n + tr(S)}{n - 2 - tr(S)} \tag{7}$$

where $\hat{\sigma}$ represents the estimated standard deviation of the error term while $tr(S)$ is the trace of the hat matrix $S$. It should be noted that the bandwidths of GWR are assumed to be the same for all influencing factors. This is a relatively strong assumption as the bandwidth of different factors (namely, the area of spatial clustering) may vary with each other. The bandwidths of different influencing factors are rarely the same on the spatial dimension. Hence, it is necessary to consider more specific variations of bandwidths for different influencing factors in spatial analysis, which is not fully concerned with the aforementioned GWR. For a specific example, the bandwidth of modeling the effect of visibility may differ from that of modeling the influences of wind. Due to the limitations of the classic GWR model, several alternatives have been utilized to further improve GWR in this aspect, such as semi-parametric geographically weighted regression (SGWR) and multi-scale geographically weighted regression (MGWR) models. SGWR roughly divides bandwidths into two broad categories (i.e., local and global) to further consider varying bandwidths. Alternatively, we leverage the multi-scale geographically weighted regression to address the aforenoted limitations. MGWR is a frontier extension methodology, which allows bandwidth differentiation for different influencing factors (Oshan et al., 2019). In contrast, MGWR proves to be a more accurate and credible model than GWR by relaxing the assumption of fixed bandwidths for all the influencing factors. The formula of MGWR is

$$y_i = \beta_0(u_i, v_i) + \sum_{j=1}^{k}\beta_{bwj}(u_i, v_i)x_{ij} + \varepsilon_i \tag{8}$$

where $\beta_{bwj}$ refers to the coefficient of the $j^{th}$ factor with selecting the optimal bandwidth of the factor also by the bi-square kernel function and the AICc mentioned in GWR (Oshan et al., 2019). Different bandwidths required by MGWR imply that each relationship between the dependent variable and a specific explanatory variable should have different spatial weighting matrices. Thus, the traditional estimator used by GWR (i.e. weighted least square) is not applicable in MGWR. MGWR embeds different computational processes to operate at different spatial scales by deriving separate bandwidths for the conditional relationships. A back-fitting algorithm proposed by Fotheringham et al. (2017) is utilized to estimate the MGWR approach, where GWR estimated coefficients are treated as initial values in the back-fitting process. All the local coefficients and the optimal bandwidths are evaluated during each iteration. Iteration terminates once the score of change from successive iterations converges to a specified threshold. The residual sum of squares (RSS) is adopted as the convergence criterion for the back-fitting algorithm, and the score of change $SOC_{RSS}$ is calculated by

$$SOC_{RSS} = \left|\frac{RSS_{new} - RSS_{old}}{RSS_{new}}\right| \tag{9}$$

where $RSS_{new}$ refers to the residual sum of squares calculated in the current step, while $RSS_{old}$ denotes the residual sum of squares calculated in the last step.

## 5. Results

*5.1 Spatial pattern analysis results*

To explore the spatial pattern of the maritime accident risks, the spatial distributions of accident frequency and accident consequence (overall reported economic loss) are modeled using Kernel Density Estimation (KDE). Fig. 3 (a) and (b) exhibit the distribution of the maritime accident frequency and consequence, respectively. For better visualization, the KDE results are divided into nine classes and the search radius is set to 0.25. The redder color indicates higher density, namely a higher frequency or consequence of maritime accidents in that water area. As seen in Fig. 3(a), coastal port waters under the jurisdiction of Xiamen-Zhangzhou, Quanzhou, Fuzhou are found to be associated with higher accident frequency, consistent with the spatial distribution of accident consequence shown in Fig. 3(b). Further, KDE results of the influencing factors in this study are presented in the appendix in case of redundancy to investigate the spatial frequency distributions of these factors. Specifically, the water area between Zhangzhou and Xiamen manifests the highest densities of collisions and fire/explosions, as shown in Fig. A1. Contact, striking rocks, grounding, and other accident types occurred more frequently around Quanzhou waters. This may be attributed to the lower water depth and the complex hydrological environment in this water. Interestingly, the area with the largest density of striking rocks is in Putian waters. Consistent with the reality, Putian coastal water area shows complex terrain with numerous reefs and rocks, which highly increase the operation difficulty of ships in avoiding hitting rocks. The water area around Fuzhou is the region with the highest density of sinking/capsizing accidents. In addition, the majority of accidents involving large ships occurred in water areas between Zhangzhou and Xiamen, as shown in Fig. A1(h).

    The results of different ship types present significantly different spatial distributions, as shown in Figs A1(i) - (l) and Fig. A1 (w). Maritime accidents involving dry cargo ships happened more frequently around Quanzhou and Fuzhou waters. Accidents involving liquid cargo ships were more frequently seen in Zhangzhou waters and Xiamen waters. Fishing ship accidents show a widespread distribution in the Fujian water area, which were mainly concentrated in southern Zhangzhou waters and the junction of Xiamen waters and Zhangzhou waters. From Figure A1(l), it is concluded that fewer accidents involving other types of ships had occurred in the Fujian water area. There was only one red area appears in the waters between Zhangzhou and Xiamen. In addition, four factors about accident cause present similar spatial. Judgment error exhibits a higher density nearby Quanzhou waters, compared with the other three accident cause factors. The spatial pattern of navigational status is similar to that of accident frequency. As to the four environmental factors, the water areas with the highest densities are Fuzhou waters for the time of the day, waters between Zhangzhou and Xiamen for visibility, Xiamen waters for the typhoon, and Quanzhou waters for strong wind/wave.

    Overall, these results show significant spatial differences in the accident consequence as well as in the potential influencing factors. More importantly, this phenomenon implies the need of investigating the influencing factors of maritime

accident frequency to reveal potential underlying reasons. It indicates that the differences in the influencing factors in different spatial dimensions may lead to the difference in accident consequence of different water areas. This motivates us to use methods such as MGWR for investigating the effects of influencing factors on maritime accident consequences.

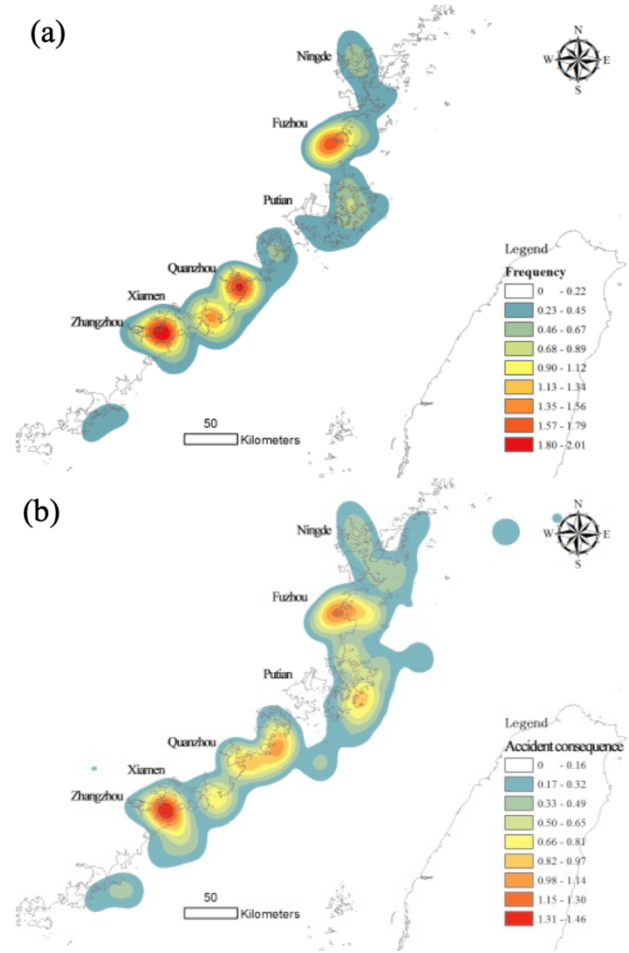

**Fig. 3.** Heat map for the maritime accidents in Fujian water area: (a) accident frequency; (b) accident consequence.

*5.2 Model comparison*

We compare the fitness performances of the adopted three regression models to find the best one. The variance inflation factors (VIFs) for each variable are checked before modeling to avoid multicollinearity issues. Finally, 19 explanatory variables with the VIFs lower than 4.5 remain in the model estimation process and are used for result interpretations, in case of multicollinearity. Table 2 presents the model performances of the three models in terms of R-Squared, AICc, log-likelihood, and RSS. As seen in Table 2, the MGWR model provides more superior performances as compared to the other two models. Specifically, the R-Squared of MGWR (0.465) is 95% and 161% higher than that of GWR (0.238) and MLR (0.178), respectively. Overall, MLR without considering the effect of spatial heterogeneity in the effects of influencing factors have poor fitness performances, indicating inefficient interpretability of MLR on our data. The MGWR model takes into consideration of the distinct effects of influencing factors in different spatial areas and also varying bandwidths on the dependent variable, and thus provides the best model fit. The results about AICs, log-likelihood, and RSS corroborate the superiority of MGWR in contrast to the MLR and GWR. Further, global Moran's I statistics of the residuals from the three models are also given in Table 2. The Moran's I index from the MGWR model is furthermore seen to be lower than that from the GWR model, suggesting that the MGWR model specification is more potent in filtering spatial autocorrelation.

Table 2. Comparison of the performance of the three models.

| Model indicators | MLR | GWR | MGWR |
| --- | --- | --- | --- |
| R-Squared | 0.178 | 0.238 | 0.465 |
| Adjusted R-Squared | 0.165 | 0.203 | 0.392 |
| AICc | 3340.391 | 3319.915 | 3102.491 |
| RSS | 1026.344 | 950.624 | 668.211 |
| Log-likelihood | -1648.818 | -1600.996 | -1381.026 |
| Moran's I in residual | 0.046 | 0.025 | 0.007 |

One advantage of MGWR as compared to the conventional GWR is that this model could accept multi-scales in geographical analysis and recognize the possible differences of spatial heterogeneity among variables simultaneously. More technically, the bandwidth in the modeling process of MGWR is varying for different variables rather than constant in spatial dimensions as the conventional GWR does. Table 3 summarizes the results about the bandwidths of variables during the modeling. It can be easily seen that all variables in GWR have the same bandwidth or spatial scale in modeling, as the conventional GWR assumes homogeneous bandwidths for all variables. Specifically, the bandwidth (i.e., spatial scale) of GWR is 792, accounting for 63.5% of the observations. In contrast, the MGWR results show significant differences in bandwidths among different variables, which range from 50 to 1246. According to the results, seven influencing factors including sinking/capsizing, liquid cargo ships, gross tonnage, time of the day, visibility, strong wind/wave, and lookout failure show quite low bandwidths, which implies that the effects of these factors on the accident consequence show noticeable distinctions in different areas. However, the other twelve factors have bandwidths of nearly 1246, which indicates the effects of these factors do not present considerable spatial heterogeneity and thus can be treated as global-scale variables in the analysis.

Table 3. Results of the three models.

| Variable | MLR | | GWR | | MGWR | |
| --- | --- | --- | --- | --- | --- | --- |
| | Coefficient | VIF | Coefficient (Mean value) | Bandwidth | Coefficient (Mean value) | Bandwidth |
| **Accident types** | | | | | | |
| Collision | 0.029 | 4.20 | 0.027 | 792 | 0.043 | 1246 |
| Grounding | -0.023 | 1.29 | -0.033 | 792 | -0.023 | 1246 |
| Striking rocks | 0.043 | 1.33 | 0.023 | 792 | 0.033 | 1246 |
| Sinking/capsizing | 0.280** | 1.23 | 0.297 | 792 | 0.274 | 50 |
| Fire/explosion | 0.055* | 1.33 | 0.048 | 792 | 0.041 | 1196 |
| **Ship information** | | | | | | |
| Two dry cargo ship involved | 0.044 | 3.46 | 0.033 | 792 | 0.013 | 1246 |
| One dry cargo ship involved | 0.010 | 2.05 | 0.017 | 792 | 0.008 | 1246 |
| Fishing ship | 0.071* | 2.15 | 0.058 | 792 | 0.039 | 1246 |
| Liquid cargo ship | 0.072** | 1.34 | 0.063 | 792 | 0.039 | 508 |
| Gross tonnage | 0.261** | 1.05 | 0.293 | 792 | 0.234 | 106 |
| Navigational status | 0.093** | 1.11 | 0.102 | 792 | 0.073 | 1246 |
| **Environmental characteristics** | | | | | | |
| Time of the day | 0.047* | 1.02 | 0.062 | 792 | 0.057 | 831 |
| Good visibility | 0.037 | 1.11 | 0.044 | 792 | 0.057 | 115 |
| Restricted visibility | -0.024 | 1.05 | -0.030 | 792 | -0.012 | 1246 |
| Strong wind/wave | 0.124** | 1.14 | 0.123 | 792 | 0.068 | 823 |
| **Accident cause factors** | | | | | | |
| Judgment error | -0.023 | 1.08 | -0.015 | 792 | -0.001 | 1246 |
| Lookout failure | 0.037 | 1.35 | 0.049 | 792 | 0.049 | 989 |
| Operation error Machinery | -0.005 | 1.37 | 0.010 | 792 | 0.002 | 1246 |
| Failure | -0.058** | 1.32 | -0.052 | 792 | -0.036 | 1246 |
| **Intercept** | 0.000 | - | 0.000 | 792 | -0.015 | 164 |
| **Observation** | | 1248 | | 1248 | | 1248 |

Note: *$0.05 \leq p \leq 0.1$, **$p \leq 0.05$

### 5.3 Spatial heterogeneities in effects of key factors

On account of the superior performances of MGWR and identified spatial heterogeneities in effects of different factors on maritime accident consequence, we will mainly discuss the results of MGWR for interpretations and analysis. The statistical description of the coefficient of each factor is summarized in Table 4. The mean value of a factor in the results of MGWR reflect the average effect of the factor, which has similar meaning as the results of MLR. However, each influencing factor presents significantly different coefficients in different water areas according to the results of MGWR, indicating that the influence of each variable varies greatly in different locations. Fig. 4 exhibits the distributions of coefficients of some factors. As shown in Fig. 4(a), the coefficient of "good visibility" (ranging from -0.2 to 1.2) has positive influences on the consequence of maritime accidents in approximately 50% of the studied areas but negative influences in other locations. However, MLR only presents a negligible and insignificant coefficient for the factor (i.e., 0.037 in Table 3), which may be attributed to the variations in the effects of the factor in different areas (i.e., spatial heterogeneity of the effect). Although sinking/capsizing and liquid cargo ships show strong positive relationships with accident consequence in MLR, the negative influences of the two variables in some locations should not be ignored as well, as shown in Fig. 4(c) and (d). Overall, models without considering the significant spatial heterogeneity in the effects of factors such as MLR only present the average effect of a factor. These flaws may result in considerable biases and thus incorrect findings. In contrast, MGWR can provide more comprehensive results with considerations of spatial heterogeneity in the effect of each factor.

Table 4. Summary statistics for MGWR parameter estimates

| Variable | Mean | STD | Min | Median | Max |
|---|---|---|---|---|---|
| **Accident types** | | | | | |
| Collision | 0.043 | 0.004 | 0.037 | 0.042 | 0.057 |
| Grounding | -0.023 | 0.004 | -0.036 | -0.022 | -0.016 |
| Striking rocks | 0.033 | 0.005 | 0.026 | 0.031 | 0.045 |
| Sinking/capsizing | 0.274 | 0.296 | -0.550 | 0.197 | 2.361 |
| Fire/explosion | 0.041 | 0.015 | 0.012 | 0.049 | 0.061 |
| **Ship information** | | | | | |
| Two dry cargo ship involved | 0.013 | 0.001 | 0.010 | 0.013 | 0.020 |
| One dry cargo ship involved | 0.008 | 0.003 | 0.002 | 0.006 | 0.016 |
| Fishing ship | 0.039 | 0.004 | 0.034 | 0.037 | 0.061 |
| Liquid cargo ship | 0.039 | 0.054 | -0.071 | 0.039 | 0.124 |
| Gross tonnage | 0.234 | 0.218 | -0.123 | 0.175 | 2.199 |
| Navigational status | 0.073 | 0.004 | 0.066 | 0.075 | 0.081 |
| **Environmental characteristics** | | | | | |
| Time of the day | 0.057 | 0.023 | -0.008 | 0.067 | 0.095 |
| Good Visibility | 0.057 | 0.260 | -0.183 | -0.024 | 1.376 |
| Restricted Visibility | -0.012 | 0.002 | -0.018 | -0.012 | -0.008 |
| Strong wind/wave | 0.076 | 0.016 | 0.055 | 0.068 | 0.139 |
| **Accident cause factors** | | | | | |
| Judgment error | -0.001 | 0.005 | -0.018 | 0.002 | 0.003 |
| Lookout failure | 0.049 | 0.017 | 0.023 | 0.063 | 0.069 |
| Operation error | 0.002 | 0.005 | -0.007 | 0.005 | 0.008 |
| Machinery failure | -0.036 | 0.002 | -0.040 | -0.037 | -0.032 |
| **Intercept** | -0.015 | 0.150 | -0.236 | -0.064 | 0.481 |

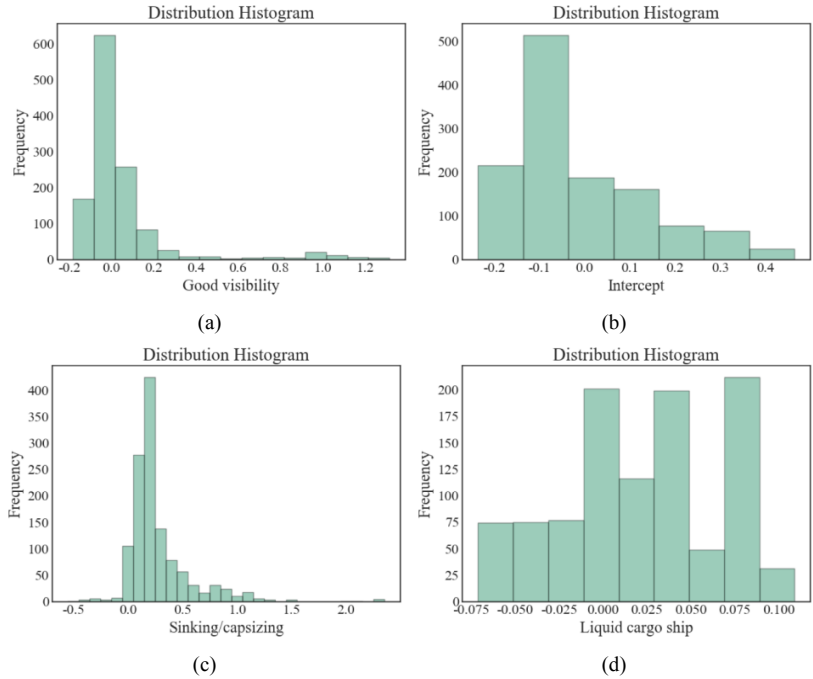

(a) (b) (c) (d)

**Fig. 4.** Coefficient distributions of influencing factors estimated by MGWR.

The following contents further discuss the influences of different factors on maritime accident consequence, and practical implications based on the results of MGWR. The parameter estimates in MGWR were mapped with natural breaks criteria, as depicted in Fig. 5 and Fig. 6. The five colors from light yellow to dark blue exhibit the values of coefficients, which deeper colors indicate a larger positive effect.

The effect of "sinking/capsizing" is positive in most locations and follows a relatively random pattern along the coastal waters, as shown in Fig. 5(b). However, accident location far from coastline normally presents a coefficient of higher than 0.370, indicating that the occurrence of sinking/capsizing accidents at these locations has much more serious consequences. Sinking/capsizing shows the highest impact in the coastal waters near Quanzhou (i.e., 0.783 - 2.361), while the accidence frequency of sinking/capsizing accidents is relatively low in this water area, as evidenced by Fig. A1(f). The results indicate that this water area has a higher probability of occurring catastrophic sinking/capsizing accidents even though the accident number is not large.

It can be seen from Fig. 5(c) that "liquid cargo ship involved" shows lower negative influences on maritime accident consequence in Putian and Quanzhou waters. However, interestingly, the large positive coefficients of "liquid cargo ship involved" are mainly distributed in coastal waters near Ningde Xiamen and Zhangzhou. This phenomenon may be attributed to the higher risk of the hazmat carried by the liquid cargo ships. Accident involving liquid cargo ships tends to associate with larger consequences (Eliopoulou and Papanikolaou, 2007). Therefore, the higher coefficients of "liquid cargo ship involved" are mainly distributed in coastal waters with higher ship density rather than in waters far from the coastline. As for "gross tonnage", negative coefficients mainly appear in southern Ningde, Fuzhou-Putian and Xiamen-Zhangzhou waters, while positive coefficients are mainly in northern Ningde, Quanzhou, and southern Zhangzhou waters, as shown in Fig. 5(d). The difference may be ascribed to the higher density of fishing ships in these water areas. The accident consequence involving fishing ships may be more sensitive to the gross tonnage of the ships due to the relatively poor safety level of fishing ships.

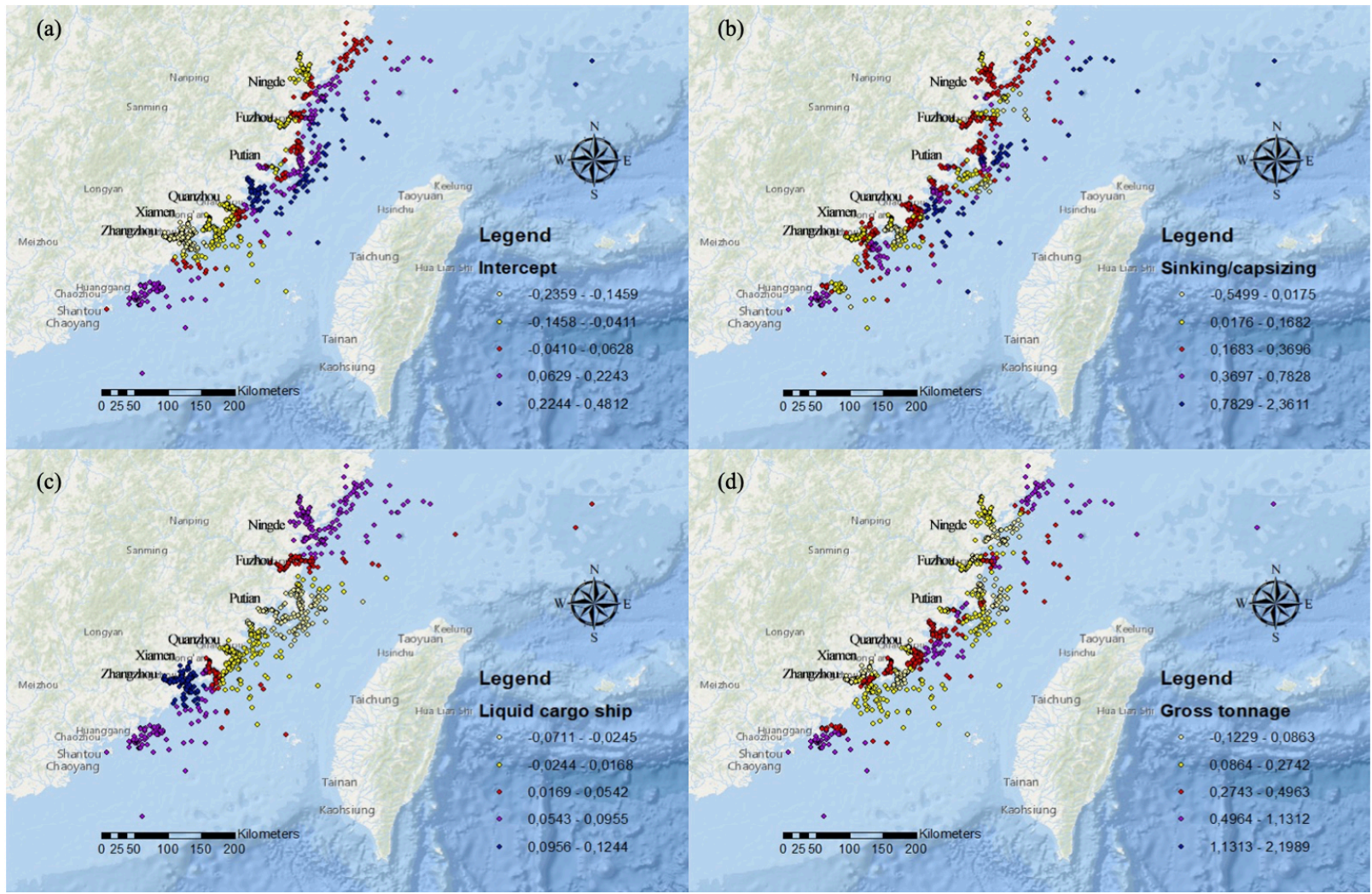

**Fig. 5.** Spatial heterogeneity in the effects of micro-scale factors on maritime accidents.

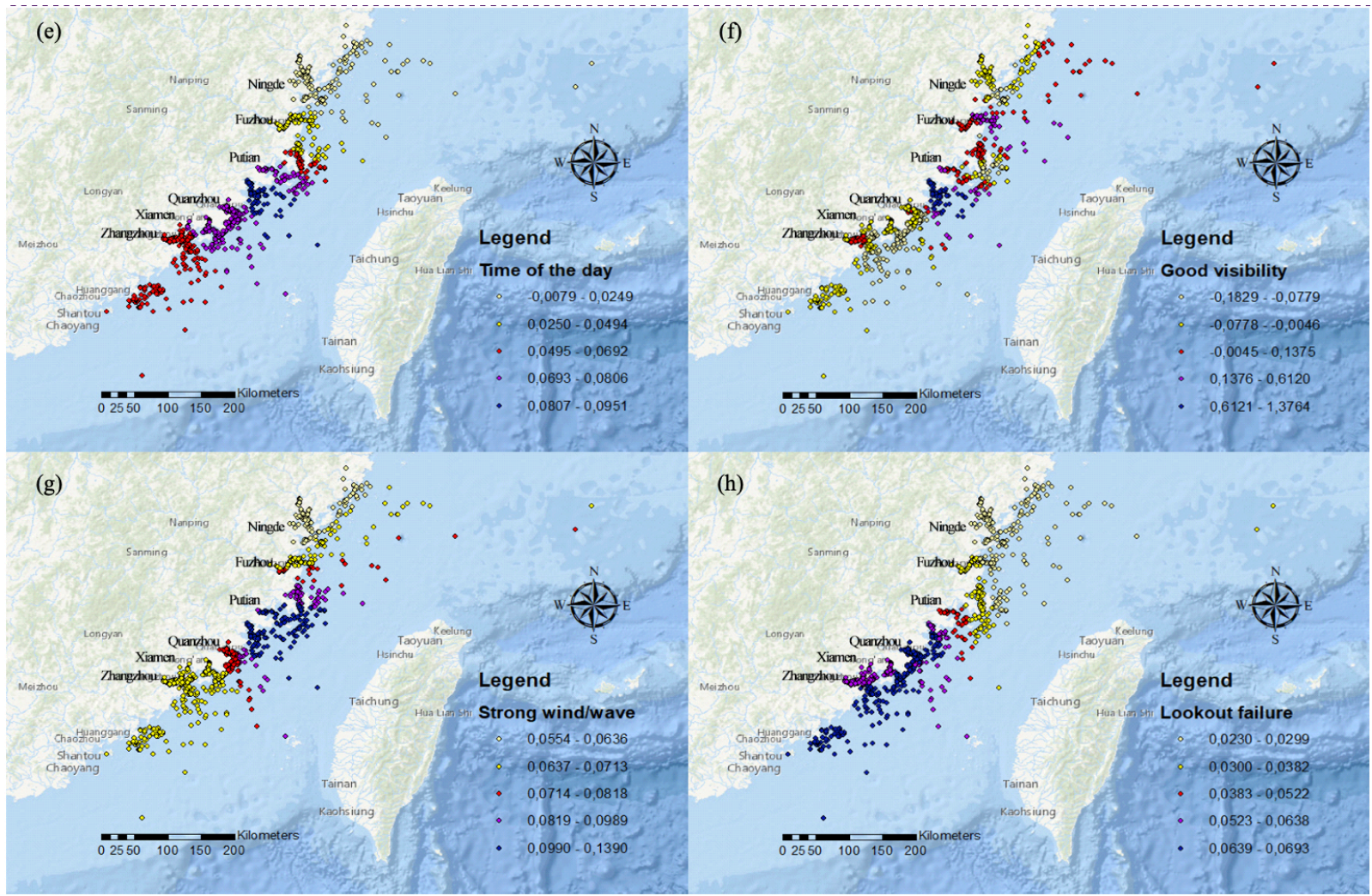

**Fig. 5.** Spatial heterogeneity in the effects of micro-scale factors on maritime accidents (Continued).

With regard to the environmental characteristics, "time of the day" from Fig. 5(e) shows larger positive coefficients in the southern Fujian waters, especially in the water area such as Putian and Quanzhou waters. This demonstrates that maritime accidents in these waters at night could lead to more serious consequences as compared to other areas. The reason may be explained by the fact that these waters are not equipped with adequate navigation marks (i.e., light buoy, light beacon and lighthouse). Furthermore, Putian waters have occurred plenty of striking rock accidents due to the complex hydrogeological environment, which could also increase the difficulty of navigating at night. From Fig. 5(f), a significant spatial heterogeneity can be observed in the coefficient of "good visibility", which shows the highest level of effects (i.e., 0.612 - 1.376) in the Putian-Quanzhou waters. Furthermore, "good visibility" also shows higher effects near Fuzhou waters (i.e., 0.138 - 0.612), while presents negative effects (i.e., -0.183 - -0.005) in Zhangzhou waters, Xiamen waters, and Ningde waters. Apart from the relatively harsh navigational conditions in Putian waters and Fuzhou waters, the higher influence of visibility may also be affected by the climate in these water areas. In all the Fujian water areas, "strong wind/wave" shows positive influences on accident consequences that are more notable those in the water area of Putian and Quanzhou, as shown in Fig. 5(g). Over the past 70 years, more than 60% of the typhoons have made landfalls pass through these waters. Therefore, the significant influences of "strong wind/waves" in the middle of the Fujian water area may be attributed to the high frequency of typhoon attacks in these waters. As shown in Fig. 5(h), "lookout failure" has positive influences in Fujian waters and is more remarkable in the south of Quanzhou waters. As mentioned before, these waters contain more fishing activities and thus result in more accidents involving fishing ships. Crews on fishing ships are normally lacking safety awareness and professional skills since these seafarers are not compulsory to have a license or other mariner credential, as evidenced by previous studies (Hovdanum et al., 2014, Li et al., 2021b).

Fig. 6 presents results about global-scale factors, namely the factors with a bandwidth of 1246. Fig. 6 (a-c) imply that "collision" and "fire/explosion" show positive coefficients on maritime accident consequence throughout the Fujian water area, while "grounding" presents slightly negative effects. Specifically, collision accidents exhibit relatively larger influences in the northern Fujian waters. Maritime accidents involving fishing ships (Fig. 6(g)) could lead to more serious consequences in Fujian waters. Moreover, the influences of "fishing ships" are more notable in the northern part of Fujian water areas. The higher influences of "fishing ships" may be explained by the fact that these water areas exist more intensive fishing activities, as partly evidenced by Fig. A1 (k). As for the accident cause factors shown in Fig. 6 (h-j), "judgment error" and "operation error" show negligible effects among the Fujian waters. Interestingly, "machinery failure" has negative influences in all the Fujian waters, which may be ascribed to the lower sailing speed after machinery failure. Furthermore, ships occurring machinery failure usually have enough rescue time to avoid serious consequences. "Navigational status" presents similar effects (i.e., 0.066 - 0.081) in the Fujian water area, as shown in Fig. 6 (k). "Restricted visibility" shows negative effects on maritime accident consequence throughout the Fujian water area. This may be due to the fact that ships are generally sailing slower under restricted visibility.

## 6. Discussions

One contribution of our study is that we found significant discrepancies in the effects of key factors of maritime accident consequences in different water areas, which could not be modelled in other traditional models. Furthermore, considering spatial scale variation between different factors makes the model results more reliable, because the influencing zones of factors on maritime accident consequences indeed varies. Aforenoted results indicate that the MGWR model is more effective and reliable when modeling the influencing factors of maritime accidents with geographical information. Specifically, compared with the traditional econometric models, MGWR model has much better modeling accuracy. This suggests that the MGWR model has a great potential to help maritime authorities accurately investigate the factors of maritime accidents in different water areas. According to the spatial heterogeneity in the effects of each influencing factor, customized safety countermeasures (or emphasis of countermeasures) can be proposed for different water areas by local maritime authorities and agencies. Specifically, with regard to reducing the probability of striking rock accidents in rocks/reef-intensive waters (i.e., in the north of Putian waters), it is suggested to set up more navigation marks and pay more attention to monitoring ship activities in these waters through closed-circuit television (CCTV) systems. Navigation marks enable ships to find out obstacles easily, especially when ships sail under poor visibility or at night. In the central Fujian waters with higher typhoon frequency, more effective wind/wave warning measures should be put in place to reduce maritime accidents caused by adverse weather conditions. For water areas with more intensive fishing activities (i.e., Zhangzhou waters and northern Ningde waters), more coast guard ships should be arranged to ensure navigational safety in these waters. Serious maritime accidents involving liquid cargo ships are normally distributed near port waters. On one hand, ports in Fujian waters should be adequately equipped with fire-fighting facilities to respond to fires/explosions caused by liquid cargo ship accidents. On the other hand, reducing the consequence of liquid cargo ship accidents should mainly focus on strengthening the emergency plan of the ship. As for waters far from the coastline, the maritime search and rescue operations should be further improved (e.g., fully mobilizing nearby ships to the accident waters for rescue) to minimize rescue response time so that can reduce the consequence of accidents occurred in these waters as much as possible. The findings thus make it possible to apply different safety strategies to different water areas with the aim of improving the safety of the specific waters. Emergency resource allocation could be optimized in accordance with economic loss prediction for different water areas based on our proposed method.

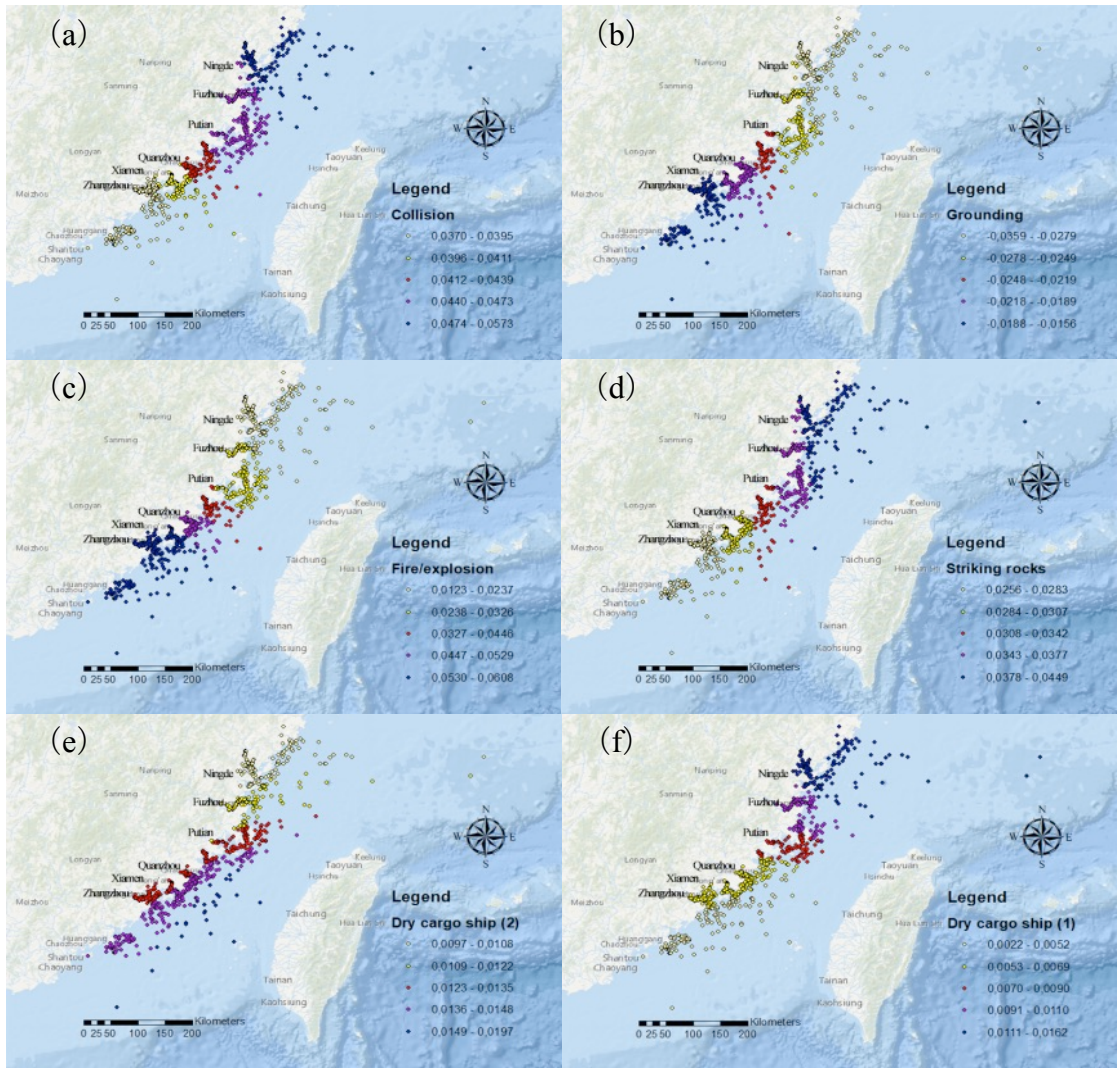

**Fig. 6.** Spatial heterogeneity in the effects of global-scale factors on maritime accident consequence.

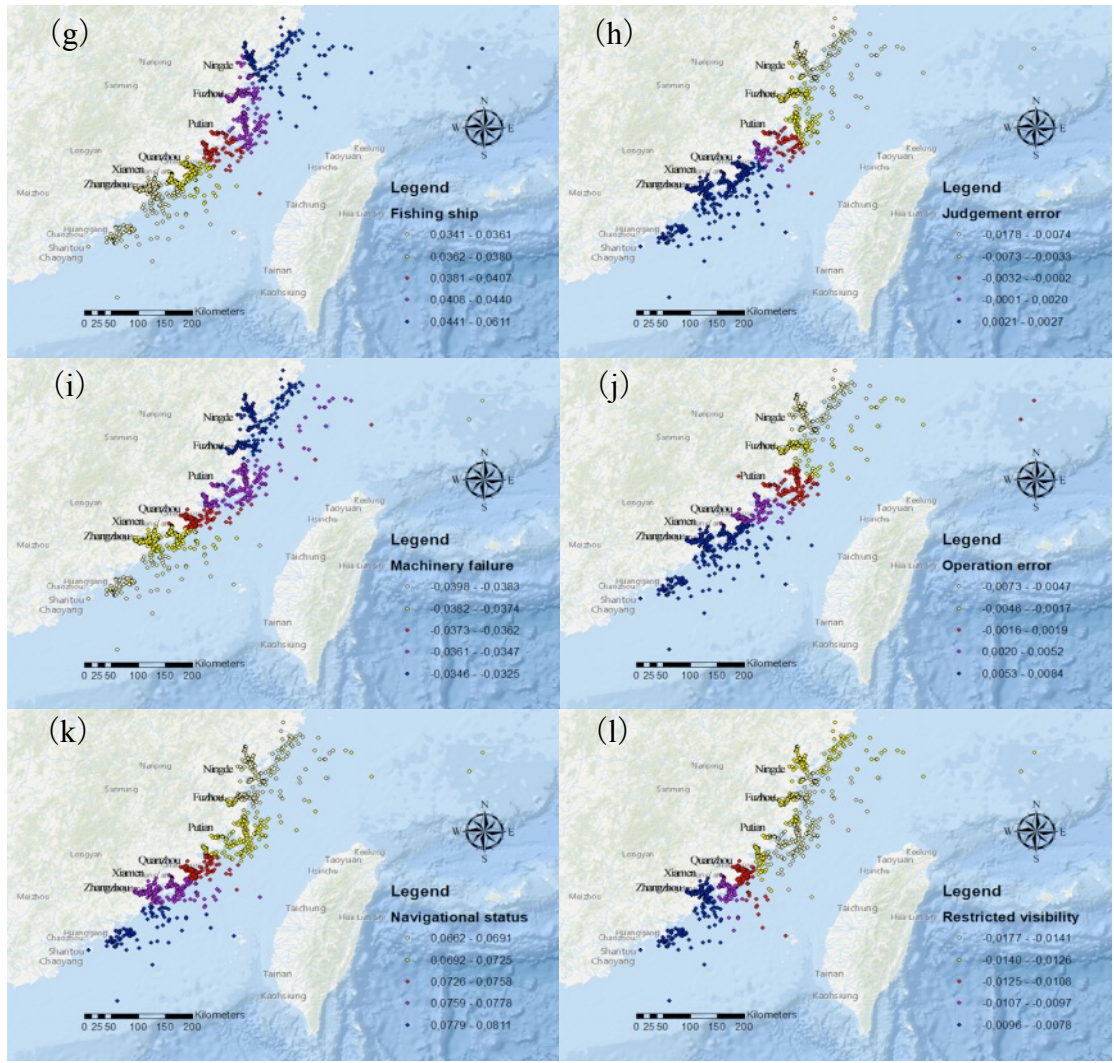

**Fig. 6.** Spatial heterogeneity in the effects of global-scale factors on maritime accident consequence (Continued).

## 7. Conclusions

The key factors of maritime accident consequences have not been comprehensively investigated, especially from a spatial perspective based on adequate maritime accident data. This study aims to fill up the gaps in relevant research by quantitatively analyzing key contributors to maritime accident consequences with special emphasis on spatial heterogeneity. Importantly, this study considers the spatial variation with multiple scales in the effects of key factors leveraging the rigorous MGWR, which is ignored by the traditional spatial econometric models such as GWR. An empirical analysis of using MGWR for analysis is presented based on maritime accident data in the East China Sea. The main findings can be summarized as follows:

- The results from Kernel Density Estimation and global Moran's I index indicate notable spatial heterogeneity in the accident sites and in the values of influencing factors. These results imply the necessity of analyzing key factors of maritime accidents consequences.
- Multi-scale geographically weighted regression presents much more fitness

superiority in modeling influences of different factors on maritime accident consequence due to its merits of considering spatial and multiple-scale variations among the variables, as compared to conventional multiple linear regression and geographically weighted regression. MGWR model can properly consider different bandwidths for different variables in spatial analysis and thus appropriately reveal the different spatial heterogeneity in the effects of a factor on maritime accident consequences.
- The results identify the key determinants of maritime accident consequences. From the global perspective, MGWR results show that sinking/capsizing, liquid cargo ship, and gross tonnage tend to be the most critical factors of all, which is consistent with the findings from MLR.
- Results reveal notably distinct influences of a factor on maritime accidence consequence in different areas, namely spatial heterogeneity. Some factors even present inverse influences in different water areas on maritime accident consequences such as sinking/capsizing, visibility, and time of the day. The results hint the necessity to take the spatial and multi-scale variations of influences of factors to properly decipher their influences on maritime accident consequences.

The findings of this study provide practical supports for maritime authorities to provide reasonable countermeasures to improve maritime safety in the specific water area. However, there are still several aspects that could be further extended. First of all, we used econometric modeling for analysis as most relevant literature did, and thus cannot directly reveal causality of the mentioned factors with maritime accident consequences. It is always a dilemma in transport safety analysis that a strong relationship does not necessarily mean causality. Further efforts are needed to investigate causality with great caution. In addition, these models are used more as a exploratory tool instead of a predictive tool. Therefore, in our future works, we will explore the possibility of using more advanced geographically weighted machine learning methods to deal with these flaws. Moreover, our case study only considers information from maritime accident data. The summary consequences and frequencies of maritime accidents in different water areas may be affected by ship traffic density, water depth, typhoon frequency, big wave frequency, fog frequency, and so on. In the future, more comprehensive information collected from AIS data, meteorological data, and hydrological data should be merged to investigate the relationship between navigational risk and various influencing factors. It is also interesting to explore the relationship between distance function (i.e., a continuous relationship of the cumulative percentages of accidents with corresponding distances to the coastline) of maritime accidents and corresponding influencing factors. Besides, our case study mainly focuses on a common water area in the East China Sea. Maritime accidents in different water areas (e.g., Arctic waters) with different meteorological and hydrological characteristics are worth investigating as well.

**Reference**

Altan, Y. and Otay, E., 2018. Spatial mapping of encounter probability in congested waterways using AIS. *Ocean Engineering*, 164, 263–271. https://doi.org/10.1016/j.oceaneng.2018.06.049

Cai, M., Zhang, J., Zhang, D., et al., 2021. Collision risk analysis on ferry ships in Jiangsu Section of the Yangtze River based on AIS data. *Reliability Engineering and System Safety*, 215, 107901. https://doi.org/10.1016/j.ress.2021.107901

Çakır, E., Fıskın, R., Sevgili, C., 2021a. Investigation of tugboat accidents severity: An application of association rule mining algorithms. *Reliability Engineering and System Safety*, 209, 107470. https://doi.org/10.1016/j.ress.2021.107470

Çakır, E., Sevgili, C., Fıskın, R., et al., 2021b. An analysis of severity of oil spill caused by vessel accidents. *Transportation Research Part D*, 90, 102662. https://doi.org/10.1016/j.trd.2020.102662

Chai, T., Xiong, D., Weng, J., et al., 2018. A Zero-Inflated Negative Binomial Regression Model to Evaluate Ship Sinking Accident Mortalities. Transportation Research Record, 2672(11), 65–72. https://doi.org/10.1177%2F0361198118787388

Chen, J., Zhang, F., Yang, C., et al., 2017. Factor and trend analysis of total-loss marine casualty using a fuzzy matter element method. *International Journal of Disaster Risk Reduction*, 24, 383–390. https://doi.org/10.1016/j.ijdrr.2017.07.001

Eliopoulou, E. and Papanikolaou, A., 2007. Casualty analysis of large tankers. *Journal of Marine Science and Technology*, 12, 240–250. https://doi.org/10.1007/s00773-007-0255-8

European Maritime Safety Agency, 2020. Annual overview of marine casualties and incidents 2020. http://emsa.europa.eu/csn-menu/items.html?cid=14&id=4266

Fotheringham, A.S., Brunsdon, C., Charlton, M., 2002. Geographically weighted regression: The analysis of spatially varying relationships. Hoboken, NJ: John Wiley & Sons.

Fotheringham, A.S. and Oshan, T.M., 2016. Geographically weighted regression and multicollinearity: Dispelling the myth. *Journal of Geographical Systems*, 18(4), 303–329. https://doi.org/10.1007/s10109-016-0239-5

Fotheringham, A.S., Yang, W., Kang, W., 2017. Multi-scale geographically weighted regression (mgwr). *Annals of the American Association of Geographers*, 107(6), 1247–1265. https://doi.org/10.1080/24694452.2017.1352480

Gao, K., Yang, Y., Li, A., et al., 2021. Spatial heterogeneity in distance decay of using bike sharing: An empirical large-scale analysis in Shanghai. *Transportation Research Part D*, 94, 102814. https://doi.org/10.1016/j.trd.2021.102814

Goerlandt, F. and Kujala, P., 2011. Traffic simulation based ship collision probability modeling. *Reliability Engineering and System Safety*, 96, 91–107. https://doi.org/10.1016/j.ress.2010.09.003

Gil, M., Kozioł, p., Wrobel, K., et al., 2022. Know your safety indicator – A determination of merchant vessels Bow Crossing Range based on big data analytics. *Reliability Engineering and System Safety*, 220, 108311. https://doi.org/10.1016/j.ress.2021.108311

Huang, D., Hu, H., Li, Y., 2013. Spatial analysis of maritime accidents using the geographic information system. *Transportation Research Record*, 2326(1), 39–44. https://doi.org/10.3141/2326-06

Hovdanum, A.S., Jensen, O.C., Petursdottir, G., Holmen, I.M., 2014. A review of fatigue in fishermen: a complicated and underprioritised area of research. *International Maritime Health*, 65 (3), 166–172. https://doi.org/10.5603/IMH.2014.0031

IMO, 2008. Casualty Investigation Code: Code of the International Standards and Recommended Practices for a Safety Investigation Into a Marine Casualty Or Marine Incident. IMO Publishing.

Jin, D., 2014. The determinants of fishing vessel accident severity. *Accident Analysis & Prevention*, 66, 1–7. https://doi.org/10.1016/j.aap.2014.01.001

Li, G., Weng, J., Fu, S., 2021a. Bootstrap-Tobit model for maritime accident economic loss considering underreporting issues. *Transportmetrica A: Transport Science*, 17(4), 1055–1076. https://doi.org/10.1080/23249935.2020.1829169

Li, G., Weng, J., Hou, Z., 2021b. Impact analysis of external factors on human errors using the ARBN method based on small-sample ship collision records. *Ocean Engineering*, 236, 109533. https://doi.org/10.1016/j.oceaneng.2021.109533

Li, S., Meng, Q., Qu, X., 2012. An overview of maritime waterway quantitative risk assessment models. *Risk Analysis*, 32, 496–512. https://doi.org/10.1111/j.1539-6924.2011.01697.x

Liu, Z., Li, Y., Zhang, Z., et al., 2022. A new evacuation accessibility analysis approach based on spatial information. *Reliability Engineering and System Safety*, 222, 108395. https://doi.org/10.1016/j.ress.2022.108395

Luo, M. and Shin, S.H., 2019. Half-century research developments in maritime accidents: Future directions. *Accident Analysis & Prevention*, 123, 448–460. https://doi.org/10.1016/j.aap.2016.04.010

Ma, X., Deng, W., Qiao, W., et al., 2022. A methodology to quantify the risk propagation of hazardous events for ship grounding accidents based on directed CN. *Reliability Engineering and System Safety*, 221, 108334. https://doi.org/10.1016/j.ress.2022.108334

Mazurek, J., Lu, L., Krata, P., et al., 2022. An updated method identifying collision-prone locations for ships. A case study for oil tankers navigating in the Gulf of Finland. *Reliability Engineering and System Safety*, 217, 108024. https://doi.org/10.1016/j.ress.2021.108024

Moran, P.A., 1950. Notes on continuous stochastic phenomena. *Biometrika*, 37 (1/2), 17–23. https://doi.org/10.2307/2332142

Mou, J., C. Tak, and H. Ligteringen. 2010. "Study on Collision Avoidance in Busy Waterways by Using AIS Data." Ocean Engineering 37: 483–490.

Murray, B., Perera, L.P., et al., 2021. An AIS-based deep learning framework for regional ship behavior prediction. *Reliability Engineering and System Safety*, 215, 107819. https://doi.org/10.1016/j.ress.2021.107819

Oshan, T.M., Li, Z., Kang, W., et al., 2019. mgwr: A Python Implementation of Multiscale Geographically Weighted Regression for Investigating Process Spatial Heterogeneity and Scale. *ISPRS International Journal of Geo-Information*, 8(6), 269. https://doi.org/10.3390/ijgi8060269

Roberts, S.E., Pettit, S.J., Marlow, P.B., 2013. Casualties and loss of life in bulk carriers from 1980 to 2010. *Maritime Policy*, 42, 223–235. https://doi.org/10.1016/j.marpol.2013.02.011

Rong, H., Teixeira, A.P., Guedes Soares, C., 2021. Spatial correlation analysis of near ship collision hotspots with local maritime traffic characteristics. *Reliability Engineering and System Safety*, 209, 107463. https://doi.org/10.1016/j.ress.2021.107463

Shi, K., Weng, J., Li, G., 2020. Exploring the effectiveness of ECA policies in reducing pollutant emissions from merchant ships in Shanghai port waters. *Marine Pollution Bulletin*, 155, 111164. https://doi.org/10.1016/j.marpolbul.2020.111164

Shi, X., Wang, Z., Li, X., et al., 2021. The effect of ride experience on changing opinions toward autonomous vehicle safety. *Communications in Transportation Research*, 1, 100003. https://doi.org/10.1016/j.commtr.2021.100003

Szlapczynski, R., Szlapczynska, J., 2021. A ship domain-based model of collision risk for near-miss detection and Collision Alert Systems. *Reliability Engineering and System Safety*, 214, 107766. https://doi.org/10.1016/j.ress.2021.107766

Talley, W.K., Jin, D., Kite-Powell, H., 2008. Determinants of the severity of cruise vessel accidents. *Transportation Research Part D*, 13, 86–94. https://doi.org/10.1016/j.trd.2007.12.001

Talley, W.K., Yip, T.L., Jin, D., 2012. Determinants of vessel-accident bunker spills. *Transportation Research Part D*, 17(8), 605–609. https://doi.org/10.1016/j.trd.2012.07.005

Tang, J., Gao, F., Liu, F., et al., 2020. Spatial heterogeneity analysis of macro-level crashes using geographically weighted Poisson quantile regression. *Accident

*Analysis & Prevention*, 148, 105833. https://doi.org/10.1016/j.aap.2020.105833Get rights and content

UNCTAD., 2020. Review of Maritime Transport 2020. United Nations publication. Sales no. E.20.II.D.31.

Wang, H., Liu, Z., Wang, X., et al., 2021a. An analysis of factors affecting the severity of marine accidents. *Reliability Engineering and System Safety*, 210, 107513. https://doi.org/10.1016/j.ress.2021.107513

Wang, L. and Yang, Z., 2018. Bayesian network modelling and analysis of accident severity in waterborne transportation: A case study in China. *Reliability Engineering & System Safety*, 180, 277–289. https://doi.org/10.1016/j.ress.2018.07.021

Wang, S., Psaraftis, N.H., Qi, J., 2021b. Paradox of international maritime organization's carbon intensity indicator. *Communications in Transportation Research*, 1, 100005. https://doi.org/10.1016/j.commtr.2021.100005

Weng, J., Yang, D., Du, G., 2018. Generalized F distribution model with random parameters for estimating property damage cost in maritime accidents. *Maritime Policy & Management*, 45(8), 963–978. https://doi.org/10.1080/03088839.2018.1475760

Xu, P. and Huang, H., 2015. Modeling crash spatial heterogeneity: random parameter versus geographically weighting. *Accident Analysis & Prevention*, 75, 16–25. https://doi.org/10.1016/j.aap.2014.10.020

Yip, T., Jin, D., Talley, W.K., 2015. Determinants of injuries in passenger vessel accidents. *Accident Analysis & Prevention*, 82, 112–117. https://doi.org/10.1016/j.aap.2015.05.025

Zhang, Y., Sun, X., Chen, J., et al., 2021. Spatial patterns and characteristics of global maritime accidents. *Reliability Engineering and System Safety*, 206, 107310. https://doi.org/10.1016/j.ress.2020.107310

Zhou, X., Chen, L., Li, M., et al., 2020. Assessing and mapping maritime transportation risk based on spatial fuzzy multi-criteria decision making: A case study in the South China sea. *Ocean Engineering*, 208, 107403. https://doi.org/10.1016/j.oceaneng.2020.107403

Ziakopoulos, A. and Yannis, G., 2020. A review of spatial approaches in road safety. *Accident Analysis & Prevention*, 135, 105323. https://doi.org/10.1016/j.aap.2019.105323

Ziakopoulos, A., 2021. Spatial analysis of harsh driving behavior events in urban networks using high-resolution smartphone and geometric data. *Accident Analysis & Prevention*, 157, 106189. https://doi.org/10.1016/j.aap.2021.106189

**Appendices**

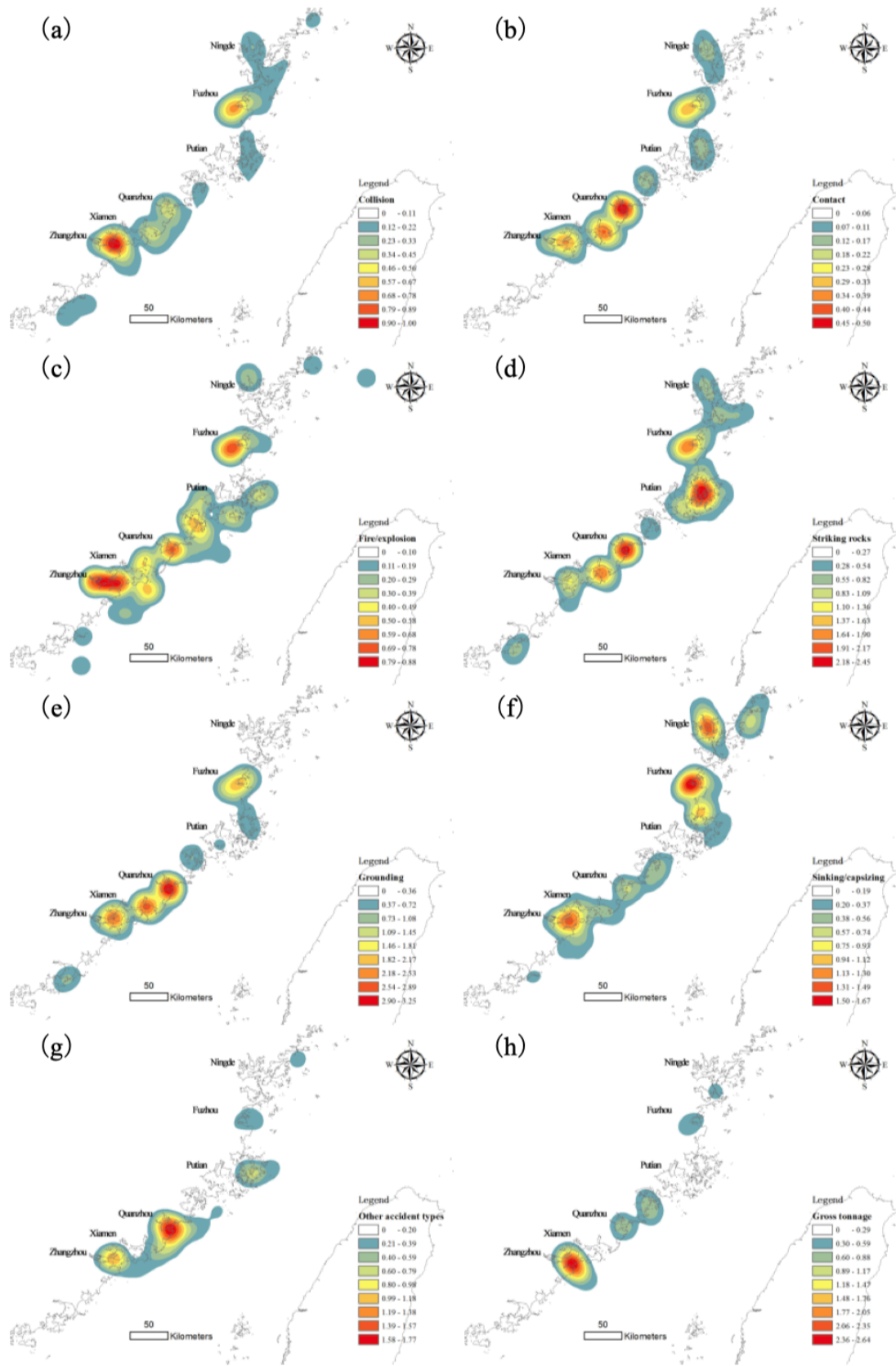

Fig. A1. Heat maps for the maritime accident variables based on KDE.

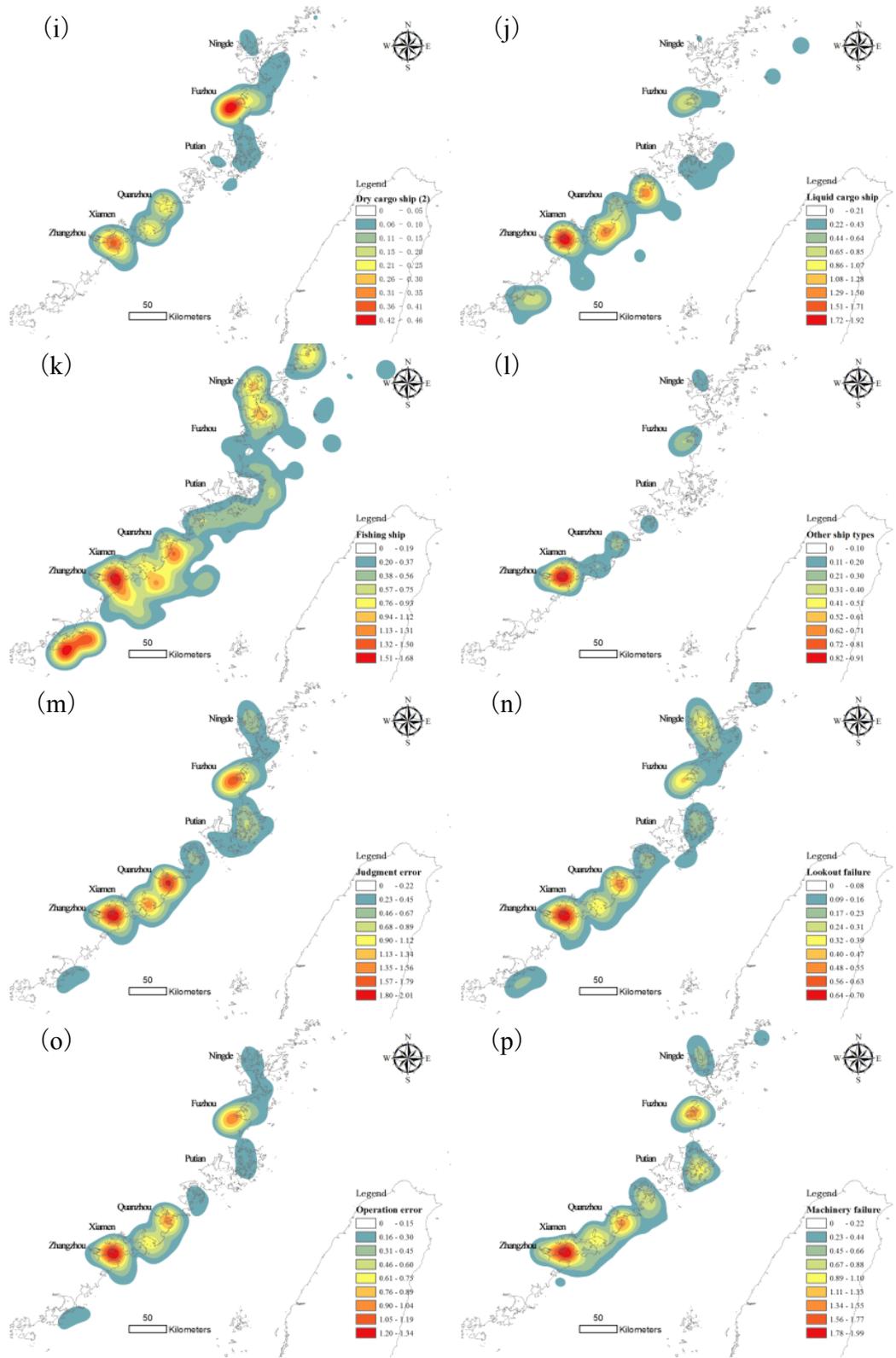

Fig. A1. (Continued).

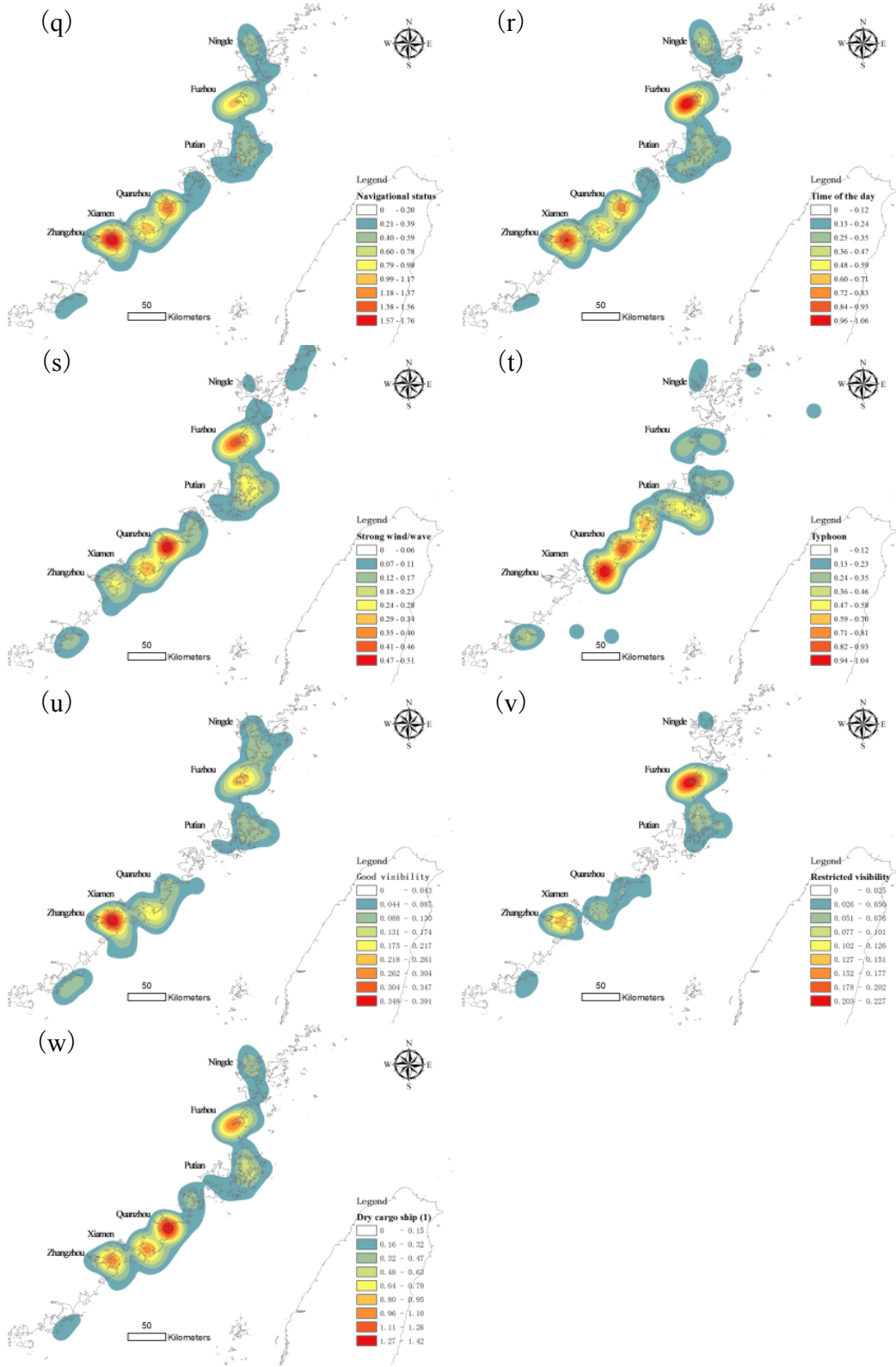

Fig. A1. (Continued)